\documentstyle[12pt]{article}
\newlength{\extraspace}
\newlength{\extraspaces}
\newcommand{\be}{\begin{equation}
\addtolength{\abovedisplayskip}{\extraspaces}
\addtolength{\belowdisplayskip}{\extraspaces}
\addtolength{\abovedisplayshortskip}{\extraspace}
\addtolength{\belowdisplayshortskip}{\extraspace}}
\newcommand{\ee}{\end{equation}}
\newcommand{\ba}{\begin{eqnarray}
\addtolength{\abovedisplayskip}{\extraspaces}
\addtolength{\belowdisplayskip}{\extraspaces}
\addtolength{\abovedisplayshortskip}{\extraspace}
\addtolength{\belowdisplayshortskip}{\extraspace}}
\newcommand{\ea}{\end{eqnarray}}
\newcommand{\nonu}{\nonumber \\[.5mm]}

\newcommand{\e}{\, {\rm e}}
\newcommand{\bra}[1]{\left\langle {#1} \right\vert}
\newcommand{\ket}[1]{\left\vert {#1} \right\rangle}

\newcommand{\Z}{\bf Z}

\newcommand{\A}{&\!\!\!}

\begin{document}
\addtolength{\baselineskip}{.7mm}
\thispagestyle{empty}
\begin{flushright}
TIT/HEP--203 \\
STUPP--92--129 \\
July, 1992
\end{flushright}
\vspace{3mm}
\begin{center}
{\large{\bf{Physical Degrees of Freedom \\[2mm]
in 2-D String Field Theories}}} \\[18mm]
{\sc Norisuke Sakai} \\[5mm]
{\it Department of Physics, Tokyo Institute of Technology \\[2mm]
Oh-okayama, Meguro, Tokyo 152, Japan} \\[7mm]
and \\[7mm]
{\sc Yoshiaki Tanii} \\[5mm]
{\it Physics Department, Saitama University \\[2mm]
Urawa, Saitama 338, Japan} \\[18mm]
{\bf Abstract}\\[7mm]
{\parbox{13cm}{\hspace{5mm}
States in the absolute (semi-relative) cohomology but not in the
relative cohomology are examined through the component
decomposition of the string field theory action for the 2-D string.
It is found that they are auxiliary fields without kinetic terms,
but are important for instance in the master equation for the
Ward-Takahashi identities. The ghost structure is analyzed in the
Siegel gauge, but it is noted that the absolute (semi-relative)
cohomology states are lost.}}
\end{center}
\vfill
\newpage
\setcounter{section}{0}
\setcounter{equation}{0}
%
%
Recent studies on matrix models and continuum Liouville approach
showed rich structures in the two-dimensional quantum gravity
coupled to conformal matter.
Especially interesting case is the two-dimensional gravity
coupled to the $c=1$ conformal matter, which can be
regarded as a string theory in two-dimensional target space and is
often called the 2-D string.
The 2-D string theory possesses infinitely many discrete states as
remnants of stringy states, and a large symmetry corresponding to the
area-preserving diffeomorphisms which
is formulated as an algebraic structure called the ground
ring \cite{WITTEN}, \cite{AVAN}, \cite{MASATA}.
It seems that the complete
informations on the 2-D string theory
may be obtained from such an algebraic approach.
In particular, it has been shown that the entire content of the
Ward-Takahashi identities from the ground ring can be summarized
as a ``master equation'' of the Batalin-Vilkovisky
type \cite{VERLINDE}.
\par
It has been shown that nontrivial cohomology elements appear for
more than one ghost numbers in the two-dimensional gravity
coupled to conformal matter \cite{LZ}.
They seem to be closely related to the rich structure
such as discrete states and the ground ring. We need to distinguish
two types of cohomologies: relative and absolute.
Absolute cohomology is defined by the states annihilated by the
BRST charge $Q_B$ but not expressed as the BRST charge acting
on other states. Relative cohomology is defined by restricting the
BRST cohomology to the subspace satisfying
\be
b_0 \ket{\Phi} = 0.
\label{relcoh}
\ee
In the case of closed string, it has been stressed that another
cohomology called semi-relative cohomology is more appropriate
instead of the absolute cohomology \cite{WIZW}.
It has recenrly been found that
the states in the semi-relative cohomology
played a key role in formulating the master equation
for the Ward-Takahashi identities \cite{VERLINDE}.
More recently, the full account of the nonpolynomial formulation
of the string field theory has appeared \cite{ZWIEBACH}.
The master equation was also formulated in this string field theory
regarding the elements of semi-relative cohomology as physical.
In spite of these important role played by the states in the
absolute cohomology or the semi-relative cohomology, only the
states in the relative cohomology are usually regarded as physical.
\par
The purpose of our note is to
examine the role played by the nontrivial elements of
the absolute cohomology in the case of open string and the
semi-relative cohomology in the case of the closed string theory.
By writing down component decomposition of the D=2 string field
theories, we find that the states in the absolute cohomology
(semi-relative cohomology) but not in the relative cohomology
have no kinetic terms and hence are auxiliary fields.
We also discuss the ghost number structure of the gauge
fixed action in the Siegel gauge, and point out that the nontrivial
elements of the absolute (semi-relative) cohomology are lost in
this particular gauge.
\par
%
%
Let us first consider the open string theory.
The BRST charge is given by
\be
Q_B = \sum_{n \in \Z} c_{-n} \left( L_n - \delta_{n, 0} \right)
- {1 \over 2} \sum_{m, n \in \Z} (m-n) : c_{-m} c_{-n} b_{m+n} :,
\label{brst}
\ee
where
\be
L_n = {1 \over 2} \sum_{m \in \Z} : \alpha^\mu_{n-m} \alpha_m^\mu :
      + {1 \over 2} (n+1) Q^\mu \alpha_n^\mu
\label{virasoro}
\ee
with $\mu = 1,\ 2$ and $Q^\mu = (0, - 2 \sqrt{2} \, i)$.
The nonvanishing (anti-)commutators of the oscillators are
\be
[ \alpha_m^\mu, \alpha_n^\nu ]
= m \delta_{m+n, \, 0} \delta^{\mu\nu}, \quad
\{ c_m, b_n \} = \delta_{m+n, \, 0}.
\label{ccr}
\ee
The normal orderings in eqs.\ (\ref{brst}) and (\ref{virasoro}) are
with respect to the Fock space vacuum $\ket{0}$ annihilated by
$\alpha_n^\mu, \ c_n, \ b_n \ (n > 0)$ and $b_0$.
We define the ghost number of the Fock vacuum to be $-{1 \over 2}$.
In order to make the BRST charge hermitian we require the
hermiticity property
\be
(\alpha_n^\mu)^\dagger
= \alpha_{-n}^\mu + Q^\mu \delta_{n, \, 0}, \quad
(c_n)^\dagger = c_{-n}, \quad
(b_n)^\dagger = b_{-n}.
\label{hermiticity}
\ee
Then we obtain $(L_n)^\dagger = L_{-n}$ and $(Q_B)^\dagger = Q_B$.
\par
To construct the gauge invariant action one has to decide
which kinds of string fields we should introduce.
Let us recall the situation in Witten's field theory of the
26-dimensional open string \cite{WITTENSFT}.
The BRST cohomology is nontrivial only for ghost numbers
$-{1 \over 2}$ and ${1 \over 2}$ (except for extra states at a
special value of 26-momentum $p^\mu=0$, which are usually
ignored) \cite{KO}.
Physical states with ghost number $-{1 \over 2}$,
which can be identified with states of the relative BRST cohomology,
represent the true nontrivial physical states.
On the other hand, physical states with ghost number ${1 \over 2}$,
which belong to the absolute BRST cohomology but not to the
relative one, are regarded as on-shell limits of pure gauge
states \cite{THORN}.
To construct the gauge invariant action a string field with
ghost number $-{1 \over 2}$ is introduced \cite{WITTENSFT}.
This single field is enough to obtain all the physical states
as solutions of the equation of motion.
\par
The string field theory of Witten type has been constructed
also for the present case of 2-D string using only a string field
with ghost number $-{1 \over 2}$ \cite{ARZU}.
However, in 2-D string, the relative BRST cohomology is nontrivial
for more than one ghost numbers \cite{LZ}.
Since the states in the relative cohomology represent dynamical
(propagating) degrees of freedom, we need to introduce at least
all the relative cohomology states in the string field theory
action. On the other hand, it is not clear whether one should
regard only states in the relative cohomology as truly physical.
Moreover, the states in the absolute (semi-relative) cohomology
play important role \cite{VERLINDE}, \cite{MMS}.
Therefore, we introduce a string field $\ket{\Phi_n(x)}$ for each
ghost number
$n = -{3 \over 2}, -{1 \over 2}, {1 \over 2}, {3 \over 2}$ for
which the absolute BRST cohomology is nontrivial.
They are states in the Fock space of the oscillators in
eq.\ (\ref{ccr}) as well as functions of the zero modes  $x^\mu$ of
the string coordinates.
To write down the action we need also a string field
$\ket{\Phi_{-{5 \over 2}}(x)}$ with ghost number $-{5 \over 2}$.
The fields $\ket{\Phi_n(x)}$ are Grassmanian odd for
$n = -{5 \over 2}, -{1 \over 2}, {3 \over 2}$ and even for
$n = -{3 \over 2}, {1 \over 2}$.
\par
The free action for these string fields is
\be
S_{\rm open} = \sum_{n=-{5 \over 2}}^{3 \over 2} \int d^2 x
\e^{i Q \cdot x} \bra{\Phi_n(x)} Q_B \ket{\Phi_{-n-1}(x)}.
\label{openaction}
\ee
The factor $\e^{i Q \cdot x}$ has been introduced so that the
operator $\alpha_0^\mu = - i \partial^\mu$ satisfies the
hermiticity condition in eq.\ (\ref{hermiticity}).
It is possible to absorb this factor into the fields:
\be
\ket{\tilde \Phi_n(x)}
= \e^{{1 \over 2}i Q \cdot x} \ket{\Phi_n(x)}.
\label{redeffield}
\ee
In terms of these redefined fields we obtain the action without
the factor $\e^{i Q \cdot x}$ but with
$\alpha_0^\mu = - i \partial^\mu$ replaced by
$\alpha_0^\mu = - i \partial^\mu - {1 \over 2} Q^\mu$.
The action (\ref{openaction}) is invariant under the gauge
transformation
\be
\delta \ket{\Phi_n(x)} = Q_B \ket{\Lambda_{n-1}(x)},
\label{gaugetrans}
\ee
where $\ket{\Lambda_{n-1}(x)}$ is an arbitrary field with ghost
number $n-1$. A set of all solutions of the equations of motion
up to the gauge transformation (\ref{gaugetrans}) coincides with
the absolute BRST cohomology.
\par
It is convenient to make the dependence on the ghost zero modes
$c_0, \ b_0$ explicit. The BRST charge and the string fields are
expanded as
\ba
Q_B \A = \A c_0 (L^{\rm tot}_0 -1) - b_0 M + \hat d, \nonu
\ket{\Phi_n(x)} \A = \A \ket{\phi_n(x)} + c_0 \ket{\psi_{n-1}(x)},
\label{zeromode}
\ea
where $L^{\rm tot}_0, \ M, \ \hat d$ do not depend on the ghost zero
modes and $b_0 \ket{\phi_{n-1}(x)} = 0 = b_0 \ket{\psi_{n-1}(x)}$.
The kinetic operator $\partial_\mu \partial_\mu$ is contained in
$L^{\rm tot}_0$.
Substituting eq.\ (\ref{zeromode}) into eq.\ (\ref{openaction})
one finds that the kinetic operator acts only on $\ket{\phi_n(x)}$.
Therefore, $\ket{\psi_{n-1}(x)}$ do not have the kinetic term and
its component fields are auxiliary fields with no independent
dynamical degrees of freedom.
\par
To see which states have independent dynamical degrees of freedom
more closely, let us consider the component field expansion up to
the first oscillator level. The expansion of the string fields are
\ba
\ket{\Phi_{-{5 \over 2}}(x)} \A = \A \cdots, \nonu
\ket{\Phi_{-{3 \over 2}}(x)} \A = \A b_{-1} \ket{0} \chi(x)
    + \cdots, \nonu
\ket{\Phi_{-{1 \over 2}}(x)} \A = \A \ket{0} \phi(x)
    + \alpha_{-1}^\mu \ket{0} i A_\mu(x)
    + c_0 b_{-1} \ket{0} \psi(x) + \cdots, \nonu
\ket{\Phi_{1 \over 2}(x)} \A = \A c_{-1} \ket{0} \psi^\ast(x)
    + c_0 \left[ \, \ket{0} \phi^\ast(x) + \alpha_{-1}^\mu \ket{0}
    i A_\mu^\ast(x) \, \right] + \cdots, \nonu
\ket{\Phi_{3 \over 2}(x)} \A = \A c_0 c_{-1} \ket{0} \chi^\ast(x)
    + \cdots,
\label{opencopm}
\ea
where the dots denote higher level states. Substituting
these expansion into the action (\ref{openaction}) we obtain
\ba
S_{\rm open}
\A = \A \int d^2 x \e^{i Q \cdot x} \biggl[ \,
{1 \over 2} \partial_\mu \phi \partial_\mu \phi - \phi^2
+ {1 \over 2} \partial_\mu A_\nu \partial_\mu A_\nu \nonu
\A\A - 2 \psi \, (\partial + i Q) \cdot A + 2 \psi \psi
- (\partial_\mu \psi^\ast + 2 A_\mu^\ast) \partial_\mu \chi
+ \cdots \, \biggr] \nonu
\A = \A \int d^2 x \e^{i Q \cdot x} \biggl[ \,
{1 \over 2} \partial_\mu \phi \partial_\mu \phi - \phi^2
+ {1 \over 4} ( \partial_\mu A_\nu - \partial_\nu A_\mu )^2 \nonu
\A\A + 2 (\psi - {1 \over 2} (\partial + i Q) \cdot A)^2
- (\partial_\mu \psi^\ast + 2 A_\mu^\ast) \partial_\mu \chi
+ \cdots \, \biggr].
\label{opencompaction}
\ea
The fields in $\ket{\psi_{n-1}(x)}$ do not have
kinetic terms and are auxiliary fields or Lagrange multiplier
fields.
Thus we find that the fields corresponding to the absolute cohomology
play the role of auxiliary fields or Lagrange multiplier fields.
The gauge transformation of the component fields are
obtained from eq.\ (\ref{gaugetrans})
\ba
\delta \chi(x) \A = \A 0, \nonu
\delta \phi(x) \A = \A 0, \nonu
\delta A_\mu(x) \A = \A - \partial_\mu \lambda(x), \nonu
\delta \psi(x) \A = \A - {1 \over 2} \partial \cdot
(\partial + i Q) \lambda(x), \nonu
\delta \psi^\ast(x) \A = \A (\partial+iQ) \cdot \epsilon(x)
- 2 \eta(x), \nonu
\delta \phi^\ast(x) \A = \A \left[ -{1 \over 2} \partial \cdot
(\partial+iQ) - 1 \right] \omega(x), \nonu
\delta A_\mu^\ast(x) \A = \A -{1 \over 2} \partial \cdot
(\partial+iQ) \epsilon_\mu(x) + \partial_\mu \eta(x), \nonu
\delta \chi^\ast(x) \A = \A {1 \over 2} \partial \cdot
(\partial+iQ) \eta^\ast(x) + (\partial+iQ) \cdot \epsilon^\ast(x),
\label{comptrans}
\ea
where fields appearing in the right hand sides are component fields
of the gauge parameter fields $\ket{\Lambda_{n-1}(x)}$,
\ba
\ket{\Lambda_{-{7 \over 2}}(x)} \A = \A \cdots, \nonu
\ket{\Lambda_{-{5 \over 2}}(x)} \A = \A \cdots, \nonu
\ket{\Lambda_{-{3 \over 2}}(x)} \A = \A b_{-1} \ket{0} \lambda(x)
+ \cdots, \nonu
\ket{\Lambda_{-{1 \over 2}}(x)} \A = \A \ket{0} \omega(x)
    + \alpha_{-1}^\mu \ket{0} i \epsilon_\mu(x)
    + c_0 b_{-1} \ket{0} \eta(x) + \cdots, \nonu
\ket{\Lambda_{1 \over 2}(x)} \A = \A c_{-1} \ket{0} \eta^\ast(x)
    + c_0 \left[ \, \ket{0} \omega^\ast(x) + \alpha_{-1}^\mu \ket{0}
    i \epsilon^\ast_\mu(x) \, \right] + \cdots.
\label{opengaugecopm}
\ea
\par
%
%
Let us discuss the gauge fixing of the action (\ref{openaction}).
The most commonly used gauge fixing is the Siegel gauge.
The gauge condition is given by
\be
b_0 \ket{\Phi_n(x)} = 0,
\label{siegel}
\ee
which is equivalent to $\ket{\psi_{n-1}(x)} = 0$
in eq.\ (\ref{zeromode}).
We would like to stress that the states in the absolute
(semi-relative) cohomology but not in the relative cohomology
are completely lost in this gauge.
Therefore one needs to invent some other gauge choice if one wants
to retain these states in the string field theory.
As an example, let us consider the fields of ghost number
$-{1 \over 2}$ at the first oscillator level, i.e. $A_\mu$ and
$\psi$. The Siegel gauge condition is $\psi = 0$.
The equations of motion derived from the gauge invariant action
(\ref{opencompaction}) have a solution in the momentum space
$\tilde A_1 = 0,\ \tilde A_2 = 1,\ \tilde \psi = -\sqrt{2}$
for $p^\mu = 0$. This solution is inconsistent with the above
gauge condition and is not allowed in the Siegel gauge.
Alternatively, we can use the Lorentz gauge
$\partial_\mu A_\mu = 0$. This gauge condition is
consistent with the above solution and all other solutions.
\par
In spite of these shortcomings, let us take for the time being the
Siegel gauge and analyze the ghost structure.
One has to introduce Faddeev-Popov ghosts, ghosts for ghosts,
ghosts for ghosts for ghosts and so on.
Perhaps the most efficient way to introduce the ghost and other
fields in the formalism is to use the Batalin-Vilkovisky
formalism \cite{BV}.
In this formalism, we should look for an solution of the ``master
equation'' containing the physical fields as leading terms.
By requiring that the states in the relative cohomology are contained
in the solution, we find that the following action is the solution
for the string field theory kinetic term
\be
S_{\rm open} = \int d^2 x
\e^{i Q \cdot x} \bra{\Phi(x)} Q_B \ket{\Phi(x)},
\label{bvsol}
\ee
where the string field $ \ket{\Phi(x)}$ is the sum over fields with
arbitrary ghost number.
The gauge fixed action can be obtained from eq.\ (\ref{bvsol})
in the same way as the 26-dimensional string \cite{THORN}.
The final result of the gauge fixed action in the Siegel gauge is
\be
S^{\rm GF}_{\rm open} = \int d^2 x
\e^{i Q \cdot x} \bra{\Phi(x)} c_0(L_0^{\rm tot}-1) \ket{\Phi(x)},
\quad b_0 \ket{\Phi(x)} = 0.
\label{gfopenaction}
\ee
\par
%
%
The above analysis for the open string theory can be repeated
for the closed string theory \cite{SIEGEL}.
The free action containing the semi-relative cohomology states is
given in terms of fields satisfying
\be
b_0^- \ket{\Psi_n(x)}=0, \quad
(L_0^{\rm tot} - \bar L_0^{\rm tot}) \ket{\Psi_n(x)}=0,
\label{closedfield}
\ee
\be
S_{\rm closed} = {1 \over 2} \int d^2 x \e^{i Q \cdot x}
\sum_{n=-4}^2 \bra{\Psi_n(x)} c_0^- (Q_B + \bar Q_B)
\ket{\Psi_{-n-2}(x)},
\label{closedaction}
\ee
where $\pm$ denote linear combinations of holomorphic and
antiholomorphic parts: $c_0^\pm = c_0 \pm \bar c_0,
\ b_0^\pm = b_0 \pm \bar b_0$.
This action is invariant under the gauge transformation which
are given for the string fields
\be
\delta \ket{\Psi_n(x)} = (Q_B + \bar Q_B) \ket{\Lambda_{n-1}(x)},
\label{closedgaugetrans}
\ee
\be
b_0^- \ket{\Lambda_{n-1}(x)}=0, \quad
(L_0^{\rm tot} - \bar L_0^{\rm tot}) \ket{\Lambda_{n-1}(x)}=0.
\label{closedgaugepara}
\ee
Since the closed string has holomorphic and antiholomorphic parts,
the ghost number of the Fock vacuum $\ket{0}$ is $-1$.
The component decomposition of the string fields is given for the
lowest nontrivial level as
\ba
\ket{\Psi_{-4}(x)}
\A = \A \cdots, \nonu
\ket{\Psi_{-3}(x)}
\A = \A b_{-1} \bar b_{-1} \ket{0} \lambda(x) + \cdots, \nonu
\ket{\Psi_{-2}(x)}
\A = \A \alpha_{-1}^\mu \bar b_{-1} \ket{0} i a_\mu(x)
+ b_{-1} \bar \alpha_{-1}^\mu \ket{0} i b_\mu(x)
+ c_0^+ b_{-1} \bar b_{-1} \ket{0} \sigma(x) + \cdots, \nonu
\ket{\Psi_{-1}(x)}
\A = \A \ket{0} \phi(x)
+ \alpha^\mu_{-1} \bar \alpha^\nu_{-1} \ket{0} h_{\mu\nu}(x)
+ b_{-1} \bar c_{-1} \ket{0} \xi(x) \nonu
\A\A - c_{-1} \bar b_{-1} \ket{0} \eta(x)
+ c_0^+ \left[ \, \alpha_{-1}^\mu \bar b_{-1} \ket{0} i A_\mu(x)
+ b_{-1} \bar \alpha_{-1}^\mu \ket{0} i B_\mu(x) \, \right]
+ \cdots, \nonu
\ket{\Psi_0(x)}
\A = \A \alpha_{-1}^\mu \bar c_{-1} \ket{0} i A^\ast_\mu(x)
+ c_{-1} \bar \alpha_{-1}^\mu \ket{0} i B^\ast_\mu(x)
+ c_0^+ \Bigl[ \, \ket{0} \phi^\ast(x) \nonu
\A\A + \alpha^\mu_{-1} \bar \alpha^\nu_{-1} \ket{0}
h^\ast_{\mu\nu}(x)
+ b_{-1} \bar c_{-1} \ket{0} \eta^\ast(x)
- c_{-1} \bar b_{-1} \ket{0} \xi^\ast(x) \, \Bigr] + \cdots, \nonu
\ket{\Psi_1(x)}
\A = \A c_{-1} \bar c_{-1} \ket{0} \sigma^\ast(x)
+ c_0^+ \left[ \, \alpha_{-1}^\mu \bar c_{-1} \ket{0}
i a^\ast_\mu(x)
+ c_{-1} \bar \alpha_{-1}^\mu \ket{0} i b^\ast_\mu(x) \, \right]
+ \cdots, \nonu
\ket{\Psi_2(x)}
\A = \A c_0^+ c_{-1} \bar c_{-1} \ket{0} \lambda^\ast(x) + \cdots.
\label{closedcomp}
\ea
Similarly to the open string case in eq.\ (\ref{comptrans}),
we obtain an component expression for the gauge transformation,
which will not be reproduced here since it is too lengthy.
In terms of the components, the action for the closed string
is given by
\ba
S_{\rm closed} \A = \A \int d^2 x \e^{i Q \cdot x}
\biggl[ \, {1 \over 2} \partial_\mu \phi \partial_\nu \phi - \phi^2
+ {1 \over 2} \partial_\lambda h_{\mu\nu} \partial_\lambda
h_{\mu\nu} - \partial_\mu \xi \partial_\mu \eta \nonu
\A\A + 2 A_\mu A_\mu + 2 B_\mu B_\mu
+ 2 A_\mu (\partial + i Q)_\nu h_{\mu\nu}
+ 2 B_\mu (\partial + i Q)_\nu h_{\nu\mu} \nonu
\A\A + 2 A_\mu \partial_\mu \xi + 2 B_\mu \partial_\mu \eta
+ \partial_\mu \sigma^\ast \partial_\mu \lambda
- 2 a^\ast_\mu \partial_\mu \lambda
+ 2 b^\ast_\mu \partial_\mu \lambda \nonu
\A\A + A^\ast_\mu \left( - \partial \cdot (\partial+iQ)a_\mu
+ 2 \partial_\mu \sigma \right)
+ B^\ast_\mu \left( - \partial \cdot (\partial+iQ)b_\mu
- 2 \partial_\mu \sigma \right) \nonu
\A\A - 2  h^\ast_{\mu\nu} ( \partial_\nu a_\mu + \partial_\mu b_\nu )
+ 2 \eta^\ast ( - (\partial + iQ) \cdot a + 2 \sigma) \nonu
\A\A - 2 \xi^\ast (  (\partial + iQ) \cdot b + 2 \sigma)
+ \cdots \, \biggr].
\label{closedcompaction}
\ea
Contributions from the string field $\ket{\Psi_{-1}(x)}$
with ghost number $-1$ can be rewritten as
\ba
S_{\rm closed}
\A = \A \int d^2 x \e^{i Q \cdot x}
\biggl[ \, {1 \over 2} \partial_\mu \phi \partial_\nu \phi - \phi^2
+ {1 \over 2} \partial_\lambda h_{\mu\nu} \partial_\lambda
h_{\mu\nu}
- {1 \over 2} \partial_\mu h_{\lambda\nu} \partial_\nu
h_{\lambda\mu} \nonu
\A\A - {1 \over 2} \partial_\mu h_{\nu\lambda} \partial_\nu
h_{\mu\lambda}
+ h_{\mu\nu} \partial_\mu \partial_\nu (\xi + \eta)
- {1 \over 2} \partial_\mu (\xi + \eta) \partial_\mu
(\xi + \eta) \nonu
\A\A + 2 (A_\mu + {1 \over 2}(\partial+iQ)_\nu h_{\mu\nu}
+ {1 \over 2} \partial_\mu \xi )^2 \nonu
\A\A + 2 (B_\mu + {1 \over 2}(\partial+iQ)_\nu h_{\nu\mu}
+ {1 \over 2} \partial_\mu \eta )^2 + \cdots \, \biggr].
\label{closedzeroghost}
\ea
Thus we see that the states in the semi-relative cohomology
but not in the relative cohomology have no kinetic terms and hence
play the role of auxiliary fields similar to
the states in the absolute cohomology in the open string theory.
The Batalin-Vilkovisky master
equation dictates that we need string field
$\ket{\Psi(x)}$ which are the sum over arbitrary ghost numbers
similarly to the open string case.
The gauge fixed action in the Siegel gauge for the closed string
is given by
\be
S^{\rm GF}_{\rm closed} = {1 \over 2} \int d^2 x
\e^{i Q \cdot x} \bra{\Psi(x)} c_0^- c_0^+
(L_0^{\rm tot} + \bar L_0^{\rm tot} -2) \ket{\Psi(x)},
\quad b_0^+ \ket{\Psi(x)} = 0.
\label{closedsiegel}
\ee
One should note that the Siegel gauge keeps all the states in the
relative cohomology but loses the states which are in the semi-relative
cohomology but are not in the relative cohomology.
\par
We thank Y. Matsumura for a collaboration in an early
stage of this work. One of us (N.S.) acknowledges Antal Jevicki
for a useful discussion. This work is supported in part by
Grant-in-Aide for Scientific Research for Priority Areas
from the Ministry of Education, Science and Culture (No. 04245211).

\end{document}